\documentclass[pra, twocolumn, showpacs, floatfix, superscriptaddress, preprintnumbers]{revtex4-1}
\usepackage{amsmath, amssymb, bbm, bbold}
\usepackage{graphicx}
\usepackage{color}

\begin{document}

\title{Incoherent ensemble dynamics in disordered systems}

\author{Clemens Gneiting}
\affiliation{Physikalisches Institut, Albert-Ludwigs-Universit\"at Freiburg, Hermann-Herder-Stra{\ss}e~3, D-79104 Freiburg, Germany}
\author{Felix~R. Anger}
\affiliation{Physikalisches Institut, Albert-Ludwigs-Universit\"at Freiburg, Hermann-Herder-Stra{\ss}e~3, D-79104 Freiburg, Germany}
\author{Andreas Buchleitner}
\affiliation{Physikalisches Institut, Albert-Ludwigs-Universit\"at Freiburg, Hermann-Herder-Stra{\ss}e~3, D-79104 Freiburg, Germany}

\date{\today}

\begin{abstract}
We derive a quantum master equation which describes the dynamics of the ensemble-averaged state of homogeneous disorder models at short times, and mediates a transition from coherent superpositions into classical mixtures. While each single realization follows unitary dynamics, this decoherence-like behavior arises as a consequence of the ensemble average. The master equation manifestly reflects the translational invariance of the disorder correlations and allows us to relate the disorder-induced dynamics to a collisional decoherence process, where the disorder correlations determine the spatial decay of coherences. We apply our theory to the (one-dimensional) Anderson model.
\end{abstract}


\preprint{\textsf{published in Phys.~Rev.~A~{93}, 032139 (2016)}}

\maketitle

\section{Introduction}

It was the insight of Anderson that disorder can substantially modify the dynamical behavior of quantum particles: The destructive quantum interference due to multiple scattering off impurities in the wire potentially brings the electrons to a halt, giving then rise to Anderson localization \cite{Anderson1958absence}. Even when the consequences of disorder are less drastic, its interplay with quantum interference can still alter the mobility pattern, causing, e.g.,~a transition from ballistic propagation to weak localization \cite{Wellens2008nonlinear}. While these interference effects already occur on the level of single disorder realizations, they even prevail under an average over many disorder realizations, this way stripping off individual peculiarities and defining a statistically robust effect. Anderson localization, e.g., unveils its characteristic trait, exponential wave function tails, on the level of the ensemble average.

The possibility to implement disorder models with highly-controllable cold atomic gases has brought it into reach to access disorder phenomena and their underlying quantum origin even on the level of the spatially resolved atomic density $n(\vec{r},t)$ \cite{Billy2008direct, Roati2008anderson}. It was, for instance, observed that the ensemble-averaged correlation function of density fluctuations exhibits, at long times, characteristic long-range correlations, which can be traced back to the macroscopic coherence in the gas \cite{Henseler2008density, Cherroret2008long}. Here, we investigate the evolution of quantum coherence under the disorder average at {\it short times}. We find that the spatial pattern of the coherence loss of the ensemble-averaged state is directly related to the correlations in the disorder potential. This loss already happens at ballistic times much shorter than the mean free time $\tau$, where the disorder does not yet have a significant effect on the level of single realizations.

We emphasize that this effective decoherence of the ensemble average state does not correspond to a loss of information as it generally occurs in the presence of an environment. In our case, single disorder realizations follow the unitary dynamics of isolated quantum systems, i.e.~the occurrence of quantum interference phenomena which survive the ensemble average, such as Anderson or weak localization, remains untouched. The coherent nature of the dynamics of single realizations can for instance be recovered by considering higher-order correlators, such as, e.g., intensity correlations. The loss of coherence of the ensemble-averaged state, on the other hand, is a consequence of the fact that different disorder realizations propagate an initially pure state into different evolved states, and that their averaging generally results in a mixed state.

To establish our results, we derive a general Lindblad master equation for the evolution of the disorder-averaged state on short time scales, allowing us to investigate the transient dynamics for arbitrary initial states $\rho_0$. In this approach, the dynamical impact of the disorder is reflected by the structure of the resulting master equation. In particular, coherent and incoherent contributions to the dynamics of the ensemble-averaged state are consistently separated. As we show, the evolution generated by the master equation for the one-dimensional (1D) Anderson model perfectly agrees with the short-time dynamics of numerically exact simulations thereof.

Let us stress that our approach lies at the interface between quantum transport theory of disordered systems and the theory of open systems. It complements other perturbative methods to treat disorder dynamics, e.g. based on averaged propagators and/or diagrammatic methods \cite{Lloyd1969exactly, Akkermans2007mesoscopic}. Alternative evolution equations for the ensemble average state have been proposed in \cite{Rammer1991quantum, Mueller2009diffusive}.

A comprehensive understanding of the disorder-induced dynamics at short times is also of practical relevance, as it permits one to access the detrimental impact of perturbations on the functioning of quantum devices. To see this, let us consider a simple example, the double-slit experiment. There, the observed fringe pattern, which represents the purpose of the device, strongly relies on the delicate interplay between the delocalized state prepared by the slits and the phases accumulated on the way to the screen. What happens if the particles are disturbed along their way, e.g.~if they propagate across a disordered scattering potential towards the screen? As we show in Fig.~1, averaging over many realizations of the disorder potential gives rise to a continuous-in-time decay of coherences, i.e.~the visibility of the interference pattern in momentum is monotonously reduced as time elapses. In other words, while single realizations exhibit distorted interference fringes, the ensemble average recovers the structure of the undisturbed pattern, however with an increasingly reduced visibility.
\begin{figure}[htb]
(a) \hspace{0.43 \columnwidth} (b) \phantom{aaaaaaaaaaaaaaaaaaa} \\
\includegraphics[width=0.48\columnwidth]{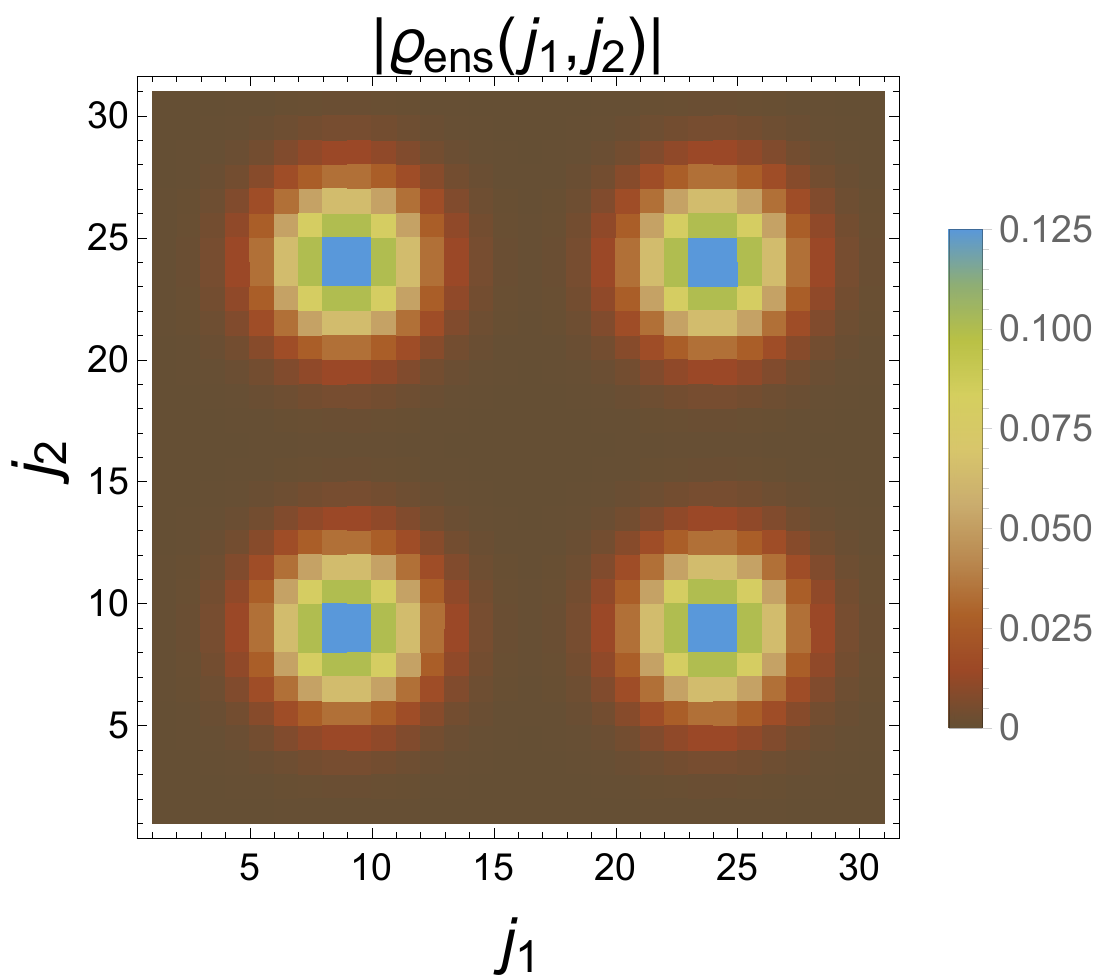}
\includegraphics[width=0.48\columnwidth]{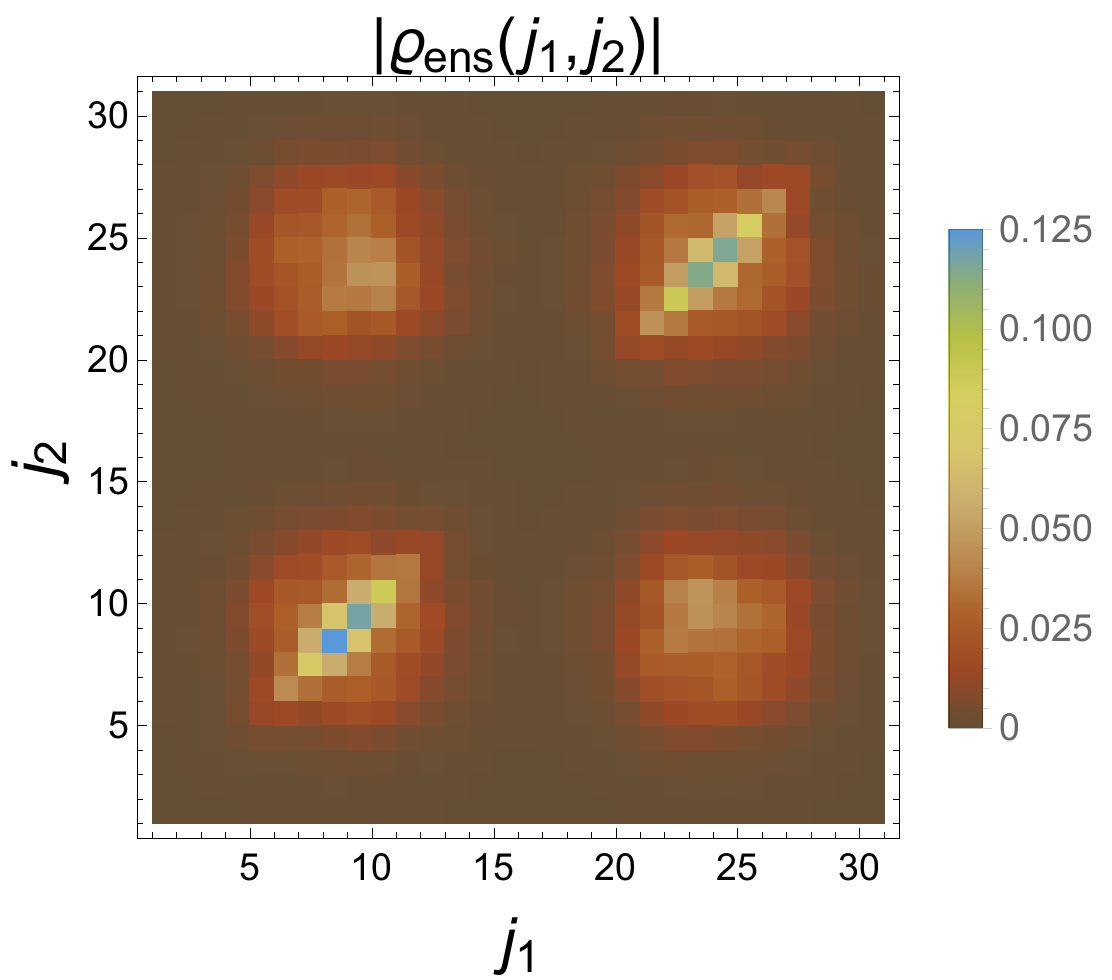}
(c) \hspace{0.43 \columnwidth} (d) \phantom{aaaaaaaaaaaaaaaaaaa} \\
\includegraphics[width=0.48\columnwidth]{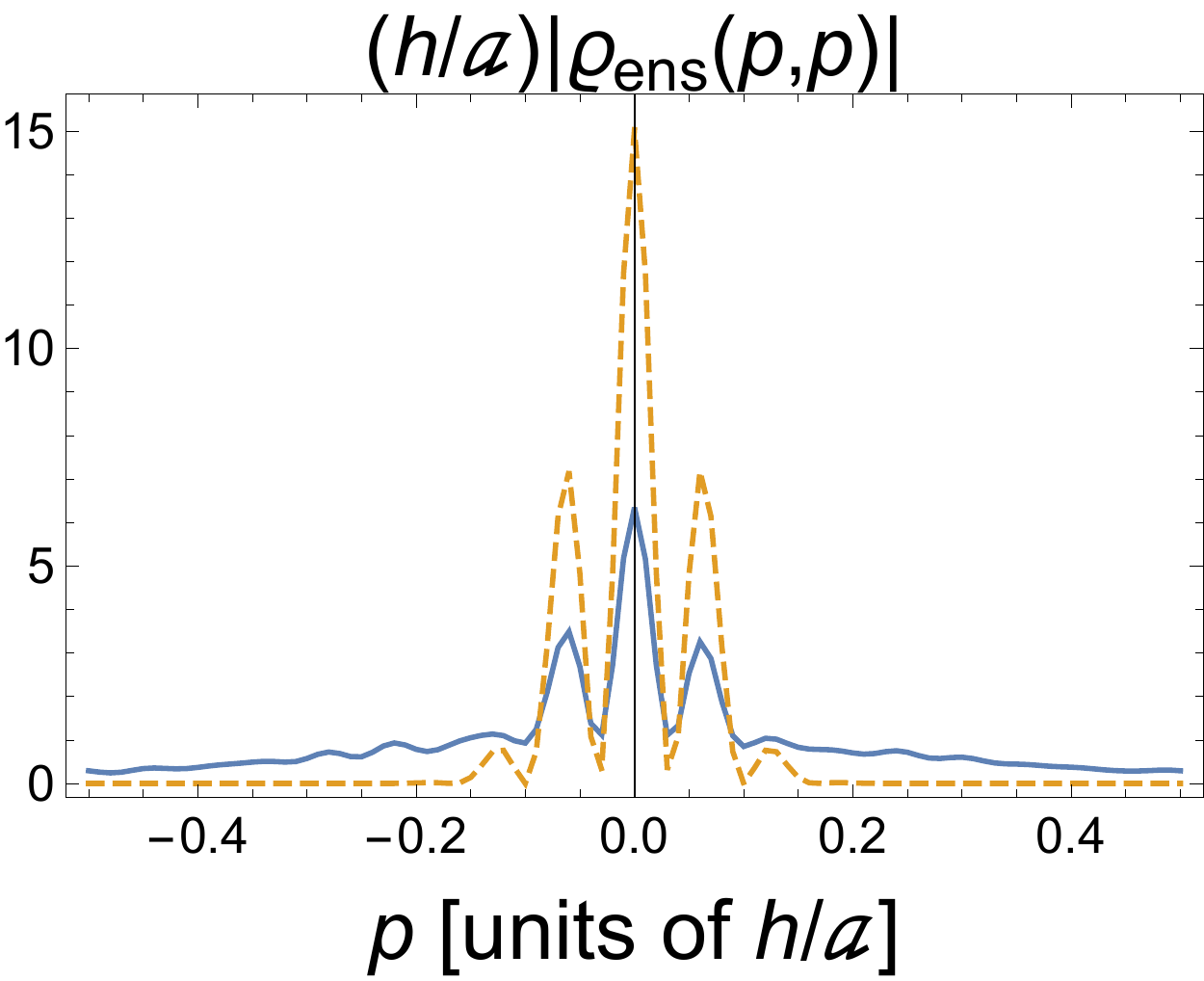}
\includegraphics[width=0.48\columnwidth]{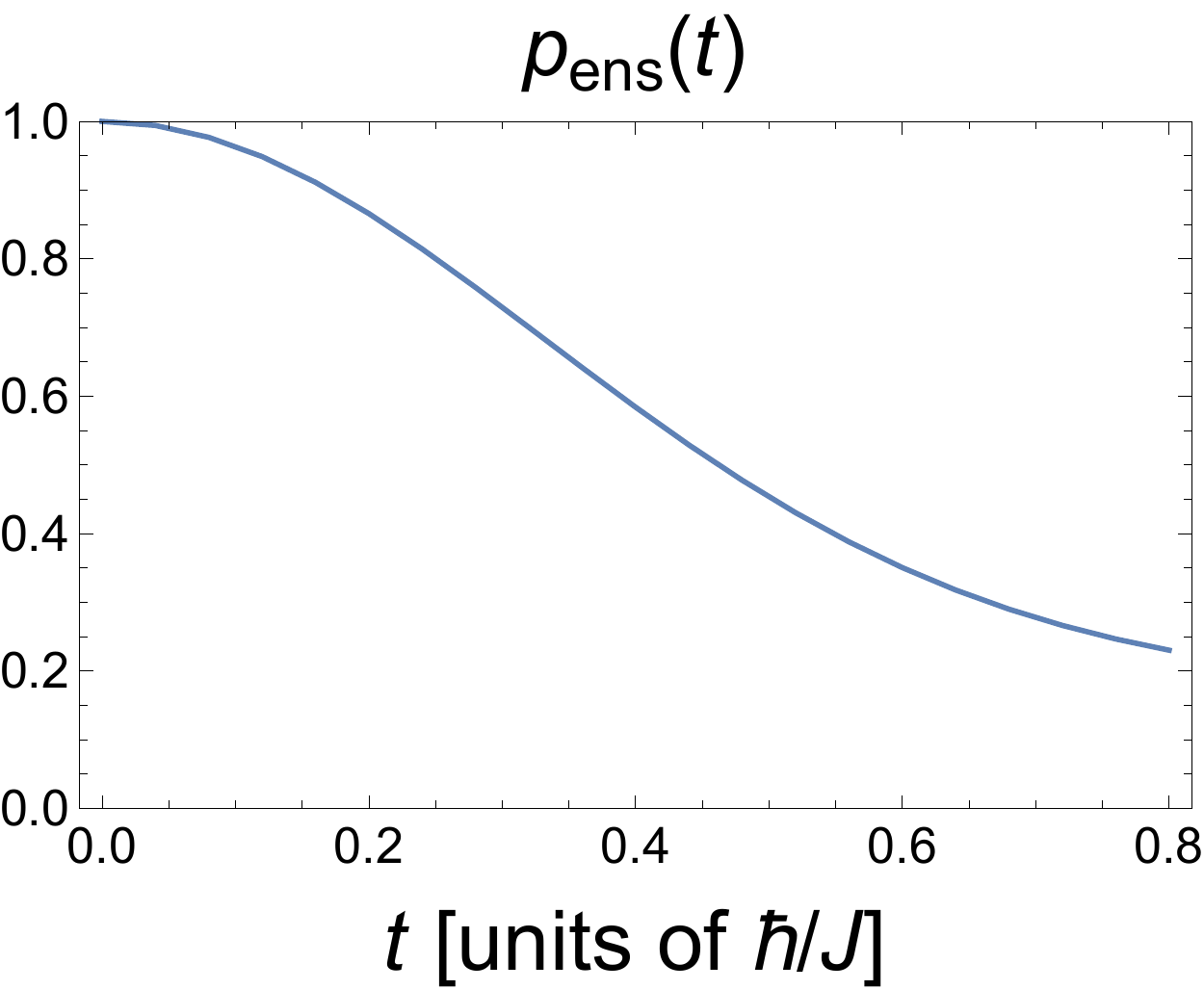}
\caption{\label{Fig:Double-slit_decoherence} (Color online) Decoherence dynamics induced by the disorder average: Evolution of an initial spatial superposition state in the (one-dimensional) Anderson model, mimicking the double-slit experiment in the presence of disorder. (a) Density matrix of the spatial superposition of two Gaussian wave packets at $t=0 \hbar/J$ (initial state). (b) Ensemble average state $\rho_{\rm ens}(t)$ at $t=0.8 \hbar/J$ ($J$ denotes the hopping constant), with disorder strength $W=5 J$, and averaged over $K=100$ realizations. One observes a decay of the coherences between the two peaks, as well as of each individual peak. The loss of coherence is also reflected in the reduced visibility of the interference pattern in momentum (c  -- blue solid at $t=0.8 \hbar/J$, orange dashed at $t=0 \hbar/J$), and in the decay of the purity $p_{\rm ens}(t)={\rm tr}[\rho_{\rm ens}(t)^2]$ of the state (d).}
\end{figure}

\section{Homogeneous disorder}

We consider a single quantum particle subject to a homogeneous disorder potential, i.e.~correlations among different locations are translationally invariant. For simplicity, we focus here on a one-dimensional, discrete (infinitely extended) configuration space, comprised of sites $|j\rangle$ with lattice spacing $a$; however, as will become clear in the course of the article, our theory works as well for continuous configuration spaces, higher dimensions, finite-size topologies, and many particles. In the case considered here, the Hamiltonian for a single disorder realization is given by
\begin{equation} \label{Eq:Anderson_model_Hamiltonian}
\hat{H}_{\vec{\varepsilon}} = - J \sum_{j \in \mathbb{Z}} (|j \rangle \langle j+1| + |j+1 \rangle \langle j|) + \sum_{j \in \mathbb{Z}} \varepsilon_j |j \rangle \langle j| ,
\end{equation}
where the (infinite-dimensional) vector $\vec{\varepsilon}$ comprised of the random on-site energies $\varepsilon_j$ distinguishes different disorder realizations. The tunneling or hopping term is characterized by the tunneling/hopping constant $J$, which controls the maximal propagation speed in the system.

The distribution $p(\vec{\varepsilon})$ of the on-site energies is assumed to be homogeneous. Besides the normalization $\left\{ \prod_{i \in \mathbb{Z}} \int {\rm d}\varepsilon_i \right\} p(\vec{\varepsilon}) = 1$, we thus require that the expectation values and two-point correlation functions satisfy
\begin{subequations} \label{Eq:Translational-invariant_distribution}
\begin{align}
&\left\{ \prod_{i \in \mathbb{Z}} \int {\rm d}\varepsilon_i \right\} p(\vec{\varepsilon}) \, \varepsilon_j = \overline{\varepsilon}_j = \overline{\varepsilon} = 0 , \label{Eq:Vanishing_expectation_values} \\
&\left\{ \prod_{i \in \mathbb{Z}} \int {\rm d}\varepsilon_i \right\} p(\vec{\varepsilon}) \, \varepsilon_j \varepsilon_{j'} = J^2 \, C(j-j') .
\end{align}
\end{subequations}
Note that we assume without loss of generality that the expectation value $\overline{\varepsilon}$ of the onsite energies vanishes. For convenience (as will become clear below), the two-point correlation function $C(\Delta j)$ is measured in units of the hopping constant $J$ (as well as all other quantities with the dimension of an energy). As we will see, it is sufficient to characterize the expectation values and the two-point correlations of the disorder distribution, as only these appear in the master-equation description at short times. Of course, homogeneity requires all higher-order correlation functions to be translationally invariant as well. We restrict ourselves to disorder which is diagonal in the site basis; however, other forms of homogeneous disorder are also conceivable.

In the case of the Anderson model \cite{Anderson1958absence}, the on-site energies of different sites are completely uncorrelated, i.e.~the disorder distribution $p(\vec{\varepsilon})$ decomposes into a product of identical single-site distributions, $p(\vec{\varepsilon}) = \prod_{i \in \mathbb{Z}} p_s(\varepsilon_i)$. The box-shaped single-site distributions $p_s(\varepsilon) = \Theta(W/2 + \varepsilon) \Theta(W/2 - \varepsilon)/W$ are characterized by the disorder strength $W$ ($\Theta$ denotes the Heaviside function). The translation-invariant correlation function is accordingly given by
\begin{equation} \label{Eq:Anderson_correlation_function}
C(j-j') = \frac{1}{12} \left( \frac{W}{J} \right)^2 \delta_{j-j',0} .
\end{equation}

\section{Short-time evolution.}

We now derive a quantum master equation which accurately describes the ensemble average dynamics of disorder models such as (\ref{Eq:Anderson_model_Hamiltonian}) at short times. In particular, it renders the, in general, incoherent nature of the ensemble average dynamics manifest in terms of the emerging Lindblad terms. To this end, we first consider the {\it unitary} time evolution for a single realization of the disorder potential, $\rho_{\vec{\varepsilon}}(t) = \hat{U}_{\vec{\varepsilon}}(t) \rho_0 \hat{U}_{\vec{\varepsilon}}^{\dagger}(t)$, with the initial state $\rho_0$ at $t_0 = 0$ and the time-evolution operator $\hat{U}_{\vec{\varepsilon}}(t) = \exp(-{\rm i} \hat{H}_{\vec{\varepsilon}} \, t/\hbar)$. Since we are interested in the evolution on short time scales, we expand to second order in ${\rm d}t$:
\begin{align} \label{Eq:Second_order_expansion_single_realization}
\rho_{\vec{\varepsilon}}({\rm d}t) =& \rho_0 + \frac{\rm i}{\hbar} {\rm d}t \, [\rho_0, \hat{H}_{\vec{\varepsilon}}] \\
&+ \frac{{\rm d}t^2}{\hbar^2} \left( \hat{H}_{\vec{\varepsilon}} \rho_0 \hat{H}_{\vec{\varepsilon}} - \frac{1}{2} \hat{H}_{\vec{\varepsilon}}^2 \rho_0 - \frac{1}{2} \rho_0 \hat{H}_{\vec{\varepsilon}}^2 \right) + \mathcal{O}({\rm d}t^3) .\nonumber
\end{align}
The second-order term is structurewise reminiscent of a Lindblad term, and, indeed, upon averaging over different realizations, the leading incoherent contributions to the disorder dynamics arise at second order in time. To see this, we take the average $\overline{\rho} = \left\{ \prod_{i \in \mathbb{Z}} \int {\rm d}\varepsilon_i \right\} p(\vec{\varepsilon}) \rho_{\vec{\varepsilon}}$ of (\ref{Eq:Second_order_expansion_single_realization}), which yields
\begin{align} \label{Eq:Second_order_expansion_ensemble_average}
\overline{\rho}({\rm d}t) = \rho_0 &+ \frac{\rm i}{\hbar} {\rm d}t \, [\rho_0, \hat{\overline{H}}] + {\rm d}t \left\{ \prod_{i \in \mathbb{Z}} \int {\rm d}\varepsilon_i \right\} \frac{p(\vec{\varepsilon}) {\rm d}t}{\hbar^2} \\
& \times \left( \hat{H}_{\vec{\varepsilon}} \rho_0 \hat{H}_{\vec{\varepsilon}} - \frac{1}{2} \hat{H}_{\vec{\varepsilon}}^2 \rho_0 - \frac{1}{2} \rho_0 \hat{H}_{\vec{\varepsilon}}^2 \right) + \mathcal{O}({\rm d}t^3) . \nonumber
\end{align}
In the first-order von Neumann term we exploited that the initial state is independent of the disorder realization. It therefore commutes with the ensemble average, resulting in the average Hamiltonian $\hat{\overline{H}} = \left\{ \prod_{i \in \mathbb{Z}} \int {\rm d}\varepsilon_i \right\} p(\vec{\varepsilon}) \hat{H}_{\vec{\varepsilon}}$. Such reduction is, in general, impossible for the second order term, which ultimately gives rise to incoherent dynamics.

To convert (\ref{Eq:Second_order_expansion_ensemble_average}) into a differential equation of Lindblad form, we must not restrict our treatment to the leading contributions in ${\rm d}t$, since we would thus lose the incoherent part of the dynamics and end with the coherent evolution induced by the average Hamiltonian $\hat{\overline{H}}$ alone. To consistently identify next-to-leading order contributions, we replace $\hat{H}_{\vec{\varepsilon}} \rightarrow (\hat{H}_{\vec{\varepsilon}}-\hat{\overline{H}}) + \hat{\overline{H}}$. Equation~(\ref{Eq:Second_order_expansion_ensemble_average}) can then be rewritten as
\begin{align} \label{Eq:Second_order_expansion_ensemble_average_consistent}
\overline{\rho}({\rm d}t) =& \rho_0 + \frac{\rm i}{\hbar} {\rm d}t \, [\rho_0, \hat{\overline{H}}] \\
&+ \frac{{\rm d}t^2}{\hbar^2} \left( \hat{\overline{H}} \rho_0 \hat{\overline{H}} - \frac{1}{2} \hat{\overline{H}}^2 \rho_0 - \frac{1}{2} \rho_0 \hat{\overline{H}}^2 \right) \nonumber \\
&+ {\rm d}t \left\{ \prod_{i \in \mathbb{Z}} \int {\rm d}\varepsilon_i \right\} \frac{p(\vec{\varepsilon}) {\rm d}t}{\hbar^2} \Big( (\hat{H}_{\vec{\varepsilon}}-\hat{\overline{H}}) \rho_0 (\hat{H}_{\vec{\varepsilon}}-\hat{\overline{H}}) \nonumber \\
&- \frac{1}{2} (\hat{H}_{\vec{\varepsilon}}-\hat{\overline{H}})^2 \rho_0 - \frac{1}{2} \rho_0 (\hat{H}_{\vec{\varepsilon}}-\hat{\overline{H}})^2 \Big) + \mathcal{O}({\rm d}t^3) , \nonumber
\end{align}
where the first two lines represent the von Neumann commutator, and the last two lines the Lindblad terms, respectively, each to second order in time. The decoherence rates associated with the Lindblad terms increase linearly in time.

It follows that Eq.~(\ref{Eq:Second_order_expansion_ensemble_average_consistent}) solves, to second order in time, a Lindblad master equation for the ensemble average state,
\begin{align} \label{Eq:Short-time_master_equation}
\dot{\overline{\rho}} =& -\frac{\rm i}{\hbar} [\hat{\overline{H}}, \overline{\rho}] + \left\{ \prod_{i \in \mathbb{Z}} \int {\rm d}\varepsilon_i \right\} \gamma_{\vec{\varepsilon}}(t) \\
& \times \left( \hat{L}_{\vec{\varepsilon}} \overline{\rho} \hat{L}_{\vec{\varepsilon}}^{\dagger} - \frac{1}{2} \hat{L}_{\vec{\varepsilon}}^{\dagger} \hat{L}_{\vec{\varepsilon}} \overline{\rho} - \frac{1}{2} \overline{\rho} \hat{L}_{\vec{\varepsilon}}^{\dagger} \hat{L}_{\vec{\varepsilon}} \right) , \nonumber
\end{align}
which captures the disorder dynamics at short times. The (time-independent) Lindblad operators $\hat{L}_{\vec{\varepsilon}}$ and the corresponding (time-dependent) decoherence rates $\gamma_{\vec{\varepsilon}}(t)$ read
\begin{equation}
\hat{L}_{\vec{\varepsilon}} = \frac{\hat{H}_{\vec{\varepsilon}} - \hat{\overline{H}}}{E_0} \qquad ; \qquad \gamma_{\vec{\varepsilon}}(t) = \frac{2 p(\vec{\varepsilon}) E_0^2}{\hbar^2} t \, ,
\end{equation}
where the characteristic energy scale $E_0$ is introduced in order to obtain the appropriate dimensions; as stated before, in the case of the model (\ref{Eq:Anderson_model_Hamiltonian}) it is conveniently chosen to be the hopping constant $J$. We thus find that the ensemble average accounts for each disorder realization by an independent Lindblad term, where the Hermitian Lindblad operators are given by the offset of the disorder Hamiltonian from the average Hamiltonian. The associated decoherence rates are proportional to the probability $p(\vec{\varepsilon})$ for the realization to occur and scale linearly in time, i.e.~the rates vanish at $t=0$. The latter expresses that there is no incoherent contribution to the dynamics at first order in time. The validity range of the short-time approximation (\ref{Eq:Short-time_master_equation}) depends on the composition of the underlying disorder ensemble and must be determined case by case. While the master equation (\ref{Eq:Short-time_master_equation}) does not require, e.g., weak disorder, the time scale on which it yields reliable predictions in general depends on the disorder strength. Below we will give a numerical estimate for the Anderson model.

We emphasize that the disorder master equation (\ref{Eq:Short-time_master_equation}) still holds for arbitrary systems and general disorder distributions, since we have not yet made use of the Hamiltonian (\ref{Eq:Anderson_model_Hamiltonian}) and/or of the homogeneous distribution (\ref{Eq:Translational-invariant_distribution}). In the Appendix, we thus evaluate the short-time disorder dynamics (\ref{Eq:Short-time_master_equation}) for two unrelated, yet instructive examples: a particle of mass $m$ in one-dimensional, continuous space, subject to a random i) linear or ii) harmonic potential. In these cases one finds that the short-time dynamics of the ensemble-averaged state  are governed by the well-known Caldeira-Leggett master equation \cite{Caldeira1983path, Diosi1993high}.

In the case of the disorder model (\ref{Eq:Anderson_model_Hamiltonian}), the average Hamiltonian is given by the discrete hopping term, $\hat{\overline{H}} = - J \sum_{j \in \mathbb{Z}} (|j \rangle \langle j+1| + |j+1 \rangle \langle j|)$ (we used (\ref{Eq:Vanishing_expectation_values})), and the Lindblad operators are given by the disorder potentials, $\hat{L}_{\vec{\varepsilon}} = \sum_{j \in \mathbb{Z}} (\varepsilon_j/J) |j \rangle \langle j|$ (with $E_0 = J$). This is already conceptually appealing, since it demonstrates that the Lindblad operators are diagonal in the site basis; moreover, Eq.~(\ref{Eq:Short-time_master_equation}) predicts (confirmed by observation) an initially quasi-free, dispersive evolution of the ensemble average state in addition to the loss of coherence. However, this representation is not yet viable from a practical point of view, in the sense that it is not amenable to transparent approximations or efficient numerical simulation. In the following, we derive an alternative representation for homogeneous disorder models (\ref{Eq:Translational-invariant_distribution}) which resolves these issues and, in addition, reveals a connection to collisional decoherence.

\section{Collisional decoherence master equation.}

To obtain an alternative representation for the short-time dynamics (\ref{Eq:Short-time_master_equation}) of the homogeneous disorder models (\ref{Eq:Translational-invariant_distribution}), we exploit that Lindblad master equations are invariant w.r.t~unitary transformations of the Lindblad operators. In our case this corresponds to a transformation from the position to the momentum basis. To this end, we perform the disorder integrals in (\ref{Eq:Short-time_master_equation}) and are left with the double sum over the sites appearing in the Lindblad operators,
\begin{align}
\dot{\overline{\rho}} = -\frac{\rm i}{\hbar} [\hat{\overline{H}}, \overline{\rho}] &+ \sum_{j,j' \in \mathbb{Z}} \frac{2 J^2 t}{\hbar^2} C(j-j') \Big(|j \rangle \langle j| \overline{\rho} |j' \rangle \langle j'| \nonumber \\
&- \frac{1}{2} |j \rangle \langle j|j' \rangle \langle j'| \overline{\rho} - \frac{1}{2} \overline{\rho} |j \rangle \langle j|j' \rangle \langle j'| \Big) ,
\end{align}
where we already made use of the translational invariance (\ref{Eq:Translational-invariant_distribution}). If we then rewrite the correlation function $C(j)$ in terms of its Fourier transform $G(q)$, $C(j) = \int_{-h/2 a}^{h/2 a} {\rm d}q \, {\rm e}^{{\rm i} q j a/\hbar} G(q)$ ($a$ denotes the lattice spacing), we obtain with $\hat{x} = \sum_{j \in \mathbb{Z}} j a |j \rangle \langle j|$
\begin{equation} \label{Eq:Collisional_decoherence_equation}
\dot{\overline{\rho}} = -\frac{\rm i}{\hbar} [\hat{\overline{H}}, \overline{\rho}] + \frac{2 J^2 t}{\hbar^2} \int\limits_{-h/2 a}^{h/2 a} {\rm d}q \, G(q) \left( {\rm e}^{{\rm i} q \hat{x}/\hbar} \overline{\rho} \, {\rm e}^{-{\rm i} q \hat{x}/\hbar} - \overline{\rho} \right) .
\end{equation}
This is our main result. We find that the ensemble average dynamics of homogeneous disorder models (\ref{Eq:Translational-invariant_distribution}) are at short times described by the discrete version of the collisional decoherence master equation. The (non-Hermitian, but unitary) Lindblad operators $\hat{L}_q = \exp(\mathrm{i} q \hat{x}/\hbar)$ describe momentum kicks, whose occurrence is weighted by the momentum transfer distribution $G(q)$. The latter follows by Fourier transform from the two-point disorder correlation function $C(j)$, $G(q) = (a/h) \sum_{j \in \mathbb{Z}} \exp(-\mathrm{i} q j a/\hbar) C(j)$ \footnote{While Eq.~(\ref{Eq:Collisional_decoherence_equation}) is completely determined by the two-point correlations, we expect that a master equation that goes beyond short times systematically includes higher-order correlation functions, as well.}.

The collisional decoherence master equation (\ref{Eq:Collisional_decoherence_equation}) is usually known from an open-system context \cite{Gallis1990environmental, Hornberger2003collisional}, where it describes the decoherence that a heavy test particle undergoes due to scattering in a background gas of light particles, i.e.~no appreciable energy exchange occurs. It represents the simplest manifestation of a translational-covariant Lindblad master equation \cite{Kossakowski1972quantum, Manita1991properties, Botvich1991translation, Holevo1995translation}.

The master equation (\ref{Eq:Collisional_decoherence_equation}) allows us to deduce the decoherence dynamics of the homogeneous disorder models (\ref{Eq:Translational-invariant_distribution}). One can best understand the spatial decoherence behavior of (\ref{Eq:Collisional_decoherence_equation}) by neglecting the coherent dynamics according to the von Neumann commutator and solving the remaining equation in the position representation. One then obtains $\langle j| \overline{\rho}(t)|j' \rangle = \exp \left( -\frac{J^2 t^2}{\hbar^2} F(j-j') \right) \langle j| \rho_0|j' \rangle$, where the localization function $F(j-j')$ (not to be confused with the exponential localization of the particle density in the Anderson model) follows from a Fourier (back-)transformation of $G(q)$ and evaluates as
\begin{equation} \label{Eq:Localization_function}
F(j-j') = \int_{-h/2 a}^{h/2 a} {\rm d}q \, G(q) - C(j-j') .
\end{equation}
We thus find that the disorder two-point correlations $C(j-j')$ directly translate into the spatial decay of coherences, in the sense that the stronger the correlation between two sites, the longer their coherence survives. This again reflects the fact that the disorder, i.e.~the deviations among different ensemble members, gives rise to the decoherence-like behavior.

In the case of the 1D Anderson model with the correlation function (\ref{Eq:Anderson_correlation_function}), one obtains a constant momentum transfer distribution, $G(q) = \frac{a W^2}{12 h J^2}$, and the localization function reads $F(j-j') = \frac{W^2}{12 J^2} (1-\delta_{j-j', 0})$, i.e., while the populations remain unaffected, all spatial coherences undergo the same decay, independent of the separation of the two respective sites, since the sites are uncorrelated.

As a second example we consider a Gaussian random potential with the correlation function $C(j-j') = \frac{\xi}{J^2} \exp \left(- \frac{(j-j')^2 a^2}{L^2} \right)$, where $\xi$ denotes the correlation strength and $L$ the correlation length. Such correlations may, for example, emerge from a collection of Gaussian scattering potentials $v(j-j_n)$ with randomly distributed scattering centers $j_n$. This then yields the momentum transfer distribution $G(q) = \frac{\sqrt{\pi} L \xi}{h J^2} \exp \left( -\frac{L^2 q^2}{4 \hbar^2} \right)$ and the localization function $F(j-j') = \frac{\xi}{J^2} \Big( 1-\exp \Big[ - \left( \frac{(j-j') a}{L} \right)^2 \Big] \Big)$, i.e.~there is a smooth, Gaussian transition into the regime of constant decoherence ($|j-j'| a \gg L$). The coherence loss at short times caused by such Gaussian disorder correlations was also investigated in \cite{Boonpan2012loss} in terms of path-integral techniques (for Gaussian initial states in the continuum and a harmonic average potential $\hat{\overline{H}}$). In our language, the authors derive  the localization function $F(x-x') = (\xi/J^2) (1-(1+2 (\frac{x-x'}{L})^2)^{-1/2})$, which coincides in the short-range region ($|x-x'| < L/2$) with our result and shows qualitatively the same behavior in the long-range region. As the short-time master equation (\ref{Eq:Short-time_master_equation}) is derived without reference to a Hilbert space basis and therefore holds over the range of all sites, we interpret the quantitative deviation in the long-range region in terms of a breakdown of the path-integral approach.

\section{Numerical comparison.}

\begin{figure}[htb]
(a) \hspace{0.43 \columnwidth} (b) \phantom{aaaaaaaaaaaaaaaaaaa} \\
\includegraphics[width=0.48\columnwidth]{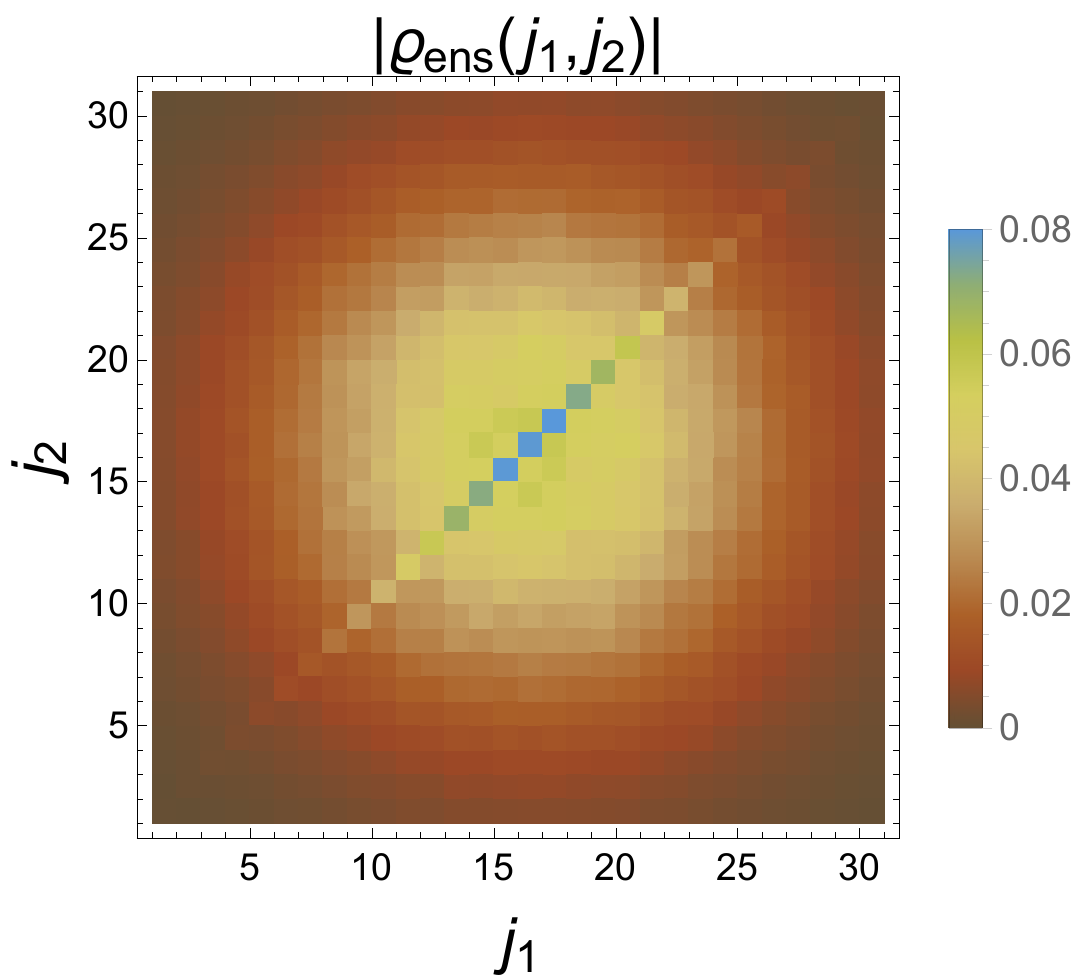}
\includegraphics[width=0.48\columnwidth]{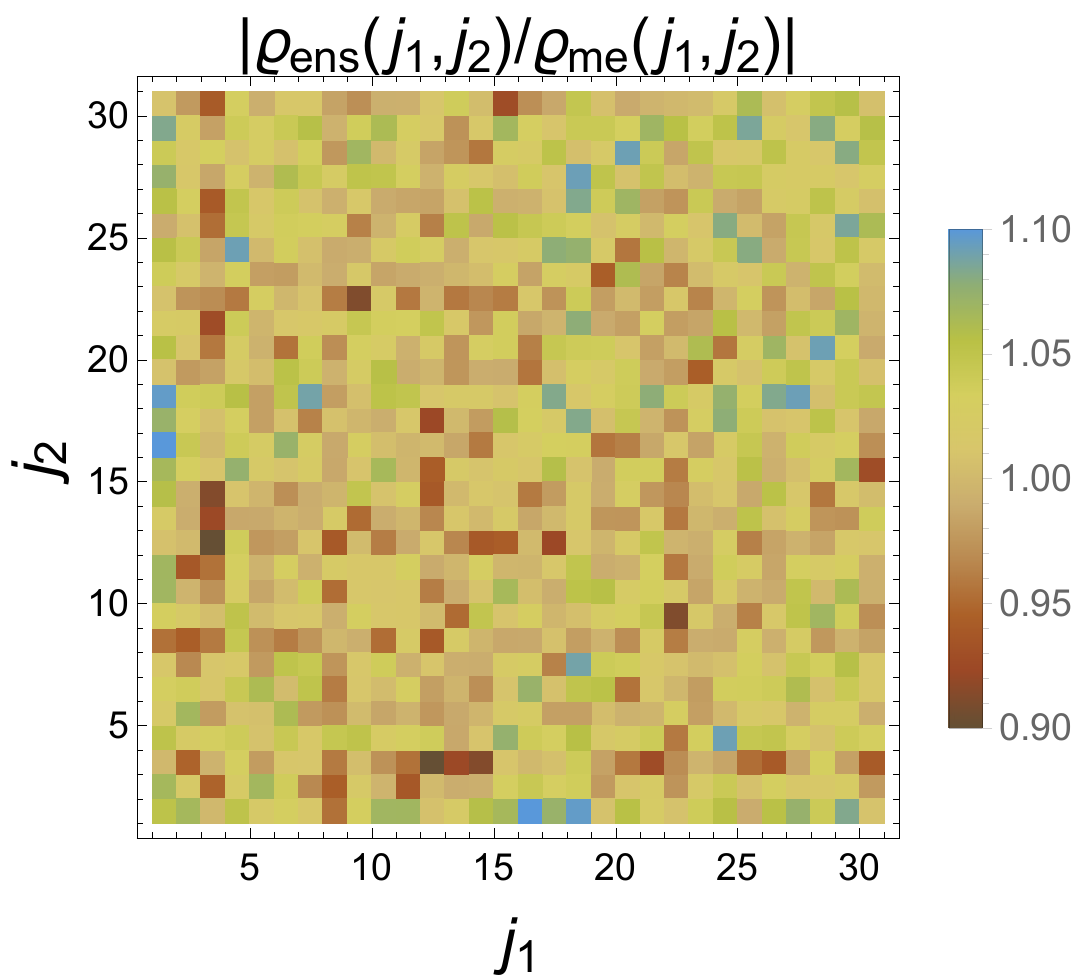}
(c) \hspace{0.43 \columnwidth} (d) \phantom{aaaaaaaaaaaaaaaaaaa} \\
\includegraphics[width=0.49\columnwidth]{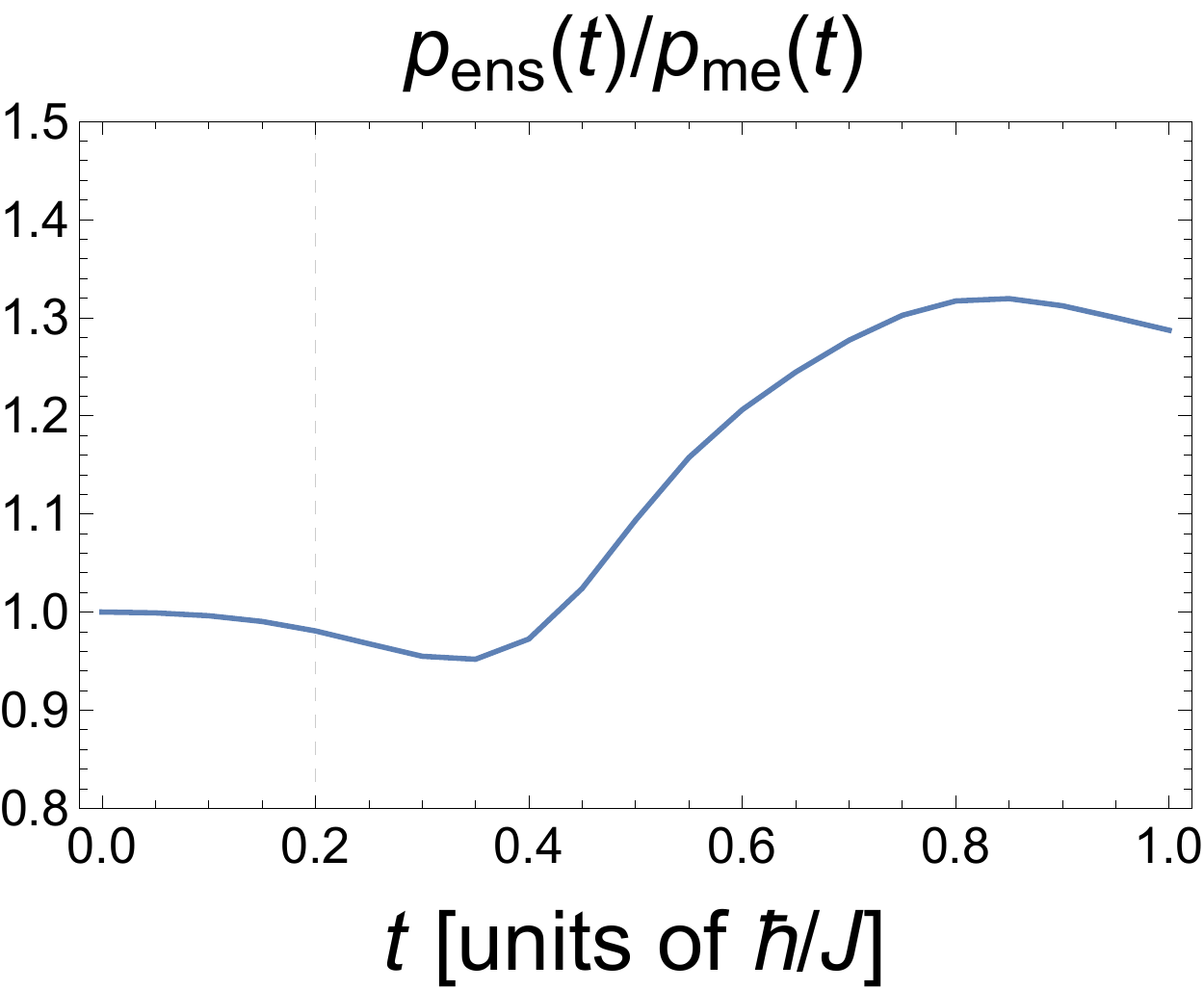}
\includegraphics[width=0.475\columnwidth]{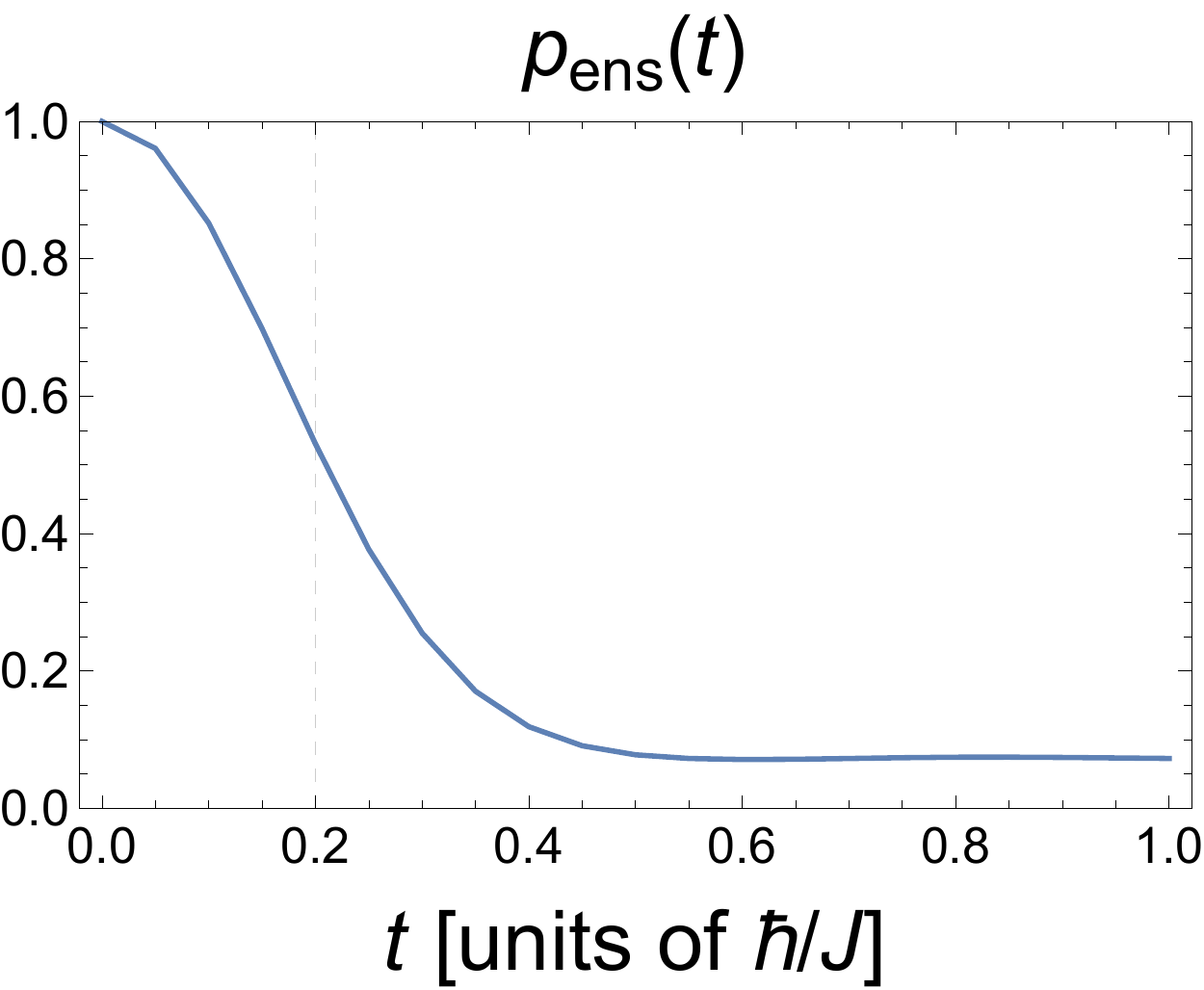}
\caption{\label{Fig:Master_equation_comparison} (Color online) Dynamics of a Gaussian initial state in the 1D Anderson model: Comparison of the time evolution $\rho_{\rm me}(t)$ predicted by the disorder master equation (\ref{Eq:Collisional_decoherence_equation}) with the numerically exact dynamics $\rho_{\rm ens}(t)$, obtained by averaging over a finite number of realizations. (a) Density matrix of the evolved ensemble average state $\rho_{\rm ens}(t)$ at $t=0.2 \hbar/J$ with strong disorder, $W=10 J$, and $K=200$ realizations. The state displays a spatially homogeneous decay of the coherences, as predicted by the master equation. (b) This is also confirmed by the ratio of the two density matrices, which remains everywhere close to one, with local fluctuations on the order of a few percent, due to the finite sampling. (c) The short-time approximation becomes poor beyond $t=0.2 \hbar/J$, which can easily be seen by inspecting the ratio of the two purities $p_{\rm ens}(t)$ and $p_{\rm me}(t)$, which starts to increasingly deviate from one. The purity provides a global measure of the decoherence and is robust, in the sense that it averages out local fluctuations due to the finite sampling. (d) At $t=0.2 \hbar/J$ the initially pure state has lost about 45\% of its purity. The latter continues to decrease monotonically and eventually converges to $p_{\rm ens} = 0.074$ beyond $t=0.5 \hbar/J$, reflecting the remaining coherence in the asymptotic state.}
\end{figure}

In order to estimate the range of validity $t_{\rm max}$ of the short-time disorder master equation (\ref{Eq:Collisional_decoherence_equation}), we compare it in case of the 1D Anderson model to the numerical ensemble average over a finite sample of disorder realizations. In Fig.~\ref{Fig:Master_equation_comparison} we show, in terms of an initial Gaussian state and for strong disorder with $W=10 J$, that the master equation correctly predicts (relative error $\pm 5\%$) the spatially homogeneous decay of the coherences up to about $t_{\rm max}=0.2 \hbar/J$, where the state has lost about $45\%$ of its initial purity $p={\rm tr}[\rho^2]$. Similarly, one obtains for disorder strengths $W=1 J$ and $W=0.1 J$ validity ranges of about $t_{\rm max}=0.9 \hbar/J$ at a purity loss of $10\%$ and $t_{\rm max}=6 \hbar/J$ at a purity loss of $1\%$, respectively. A more detailed analysis confirms that $t_{\rm max}$ roughly scales inversely with $W$, $t_{\rm max} \propto 1/W$, or, in terms of the mean free path $\ell$, $t_{\rm max} \propto \sqrt{\ell}$ (similarly, the momentum-independent decoherence rate $\gamma(t)$ scales inversely with the mean free time $\tau$, $\gamma(t) \propto t/\tau$). This suggests to interpret our theory in terms of an expansion in $W t$. Notwithstanding, we can probe the regime of strong decoherence as induced by large $W$. Let us also emphasize that we could have chosen any initial state for this analysis.

\section{Experimental verification.}

Besides the conceptual insight provided by our theory into the incoherent ensemble average dynamics of disordered quantum systems, direct experimental verifications thereof are conceivable, for example based on experiments with ultracold atoms subject to optical speckle potentials. These systems have already been successfully employed to probe the Anderson localization in the asymptotic time regime \cite{Billy2008direct, Roati2008anderson}. Moreover, it is possible to imprint various homogeneous disorder distributions on the speckle potential \cite{Deissler2010delocalization}. The restriction to short times would be implemented by simply switching the speckle potential off after the desired exposure time. Time-of-flight measurements then reveal the momentum distribution of the state. Producing an initial spatial superposition state, one may in this way observe the disorder-induced transition from a superposition into a mixture in terms of the loss of visibility of the interference pattern in momentum.

\section{Conclusions.}

We developed a theory which describes the ensemble average dynamics of disordered quantum systems at short times in terms of Lindblad master equations, with the statistical properties of the disorder potential encoded in the Lindblad terms. While this effective evolution equation accurately captures the onset of the disorder-induced coherence loss of the ensemble-averaged state in the 1D Anderson model, our theory is not yet capable to explain other disorder effects such as diffusive propagation or localization. However, a (translation-covariant) master equation which also captures the ensemble average dynamics of such disorder-induced phenomena must, in principle, exist. These must then emerge as a feature of the, in general, incoherent evolution of the ensemble-averaged state. Indeed, Fig.~\ref{Fig:Master_equation_comparison}(d) illustrates the monotonic decay of the averaged state's purity towards that of the (localized) asymptotic state. The asymptotic value of the purity decreases with increasing disorder strength $W$, which reflects that the remaining coherence in the asymptotic state is related to the localization length $\xi \propto 1/W^2$ \cite{Roemer2004weak}.

Our results represent a first step towards a treatment of disordered quantum systems in terms of quantum master equations. The impact of spectral and of unitarily invariant disorder on the dynamics of the ensemble-averaged state of finite-dimensional quantum systems at arbitrary times $t$ is the subject of \cite{Kropf2015effective}.

\section{Acknowledgments.}

C.G. thanks Valentin Volchkov for insightful discussions on the state of the art of disorder experiments with ultracold atoms. A.B. acknowledges financial support from the EU Collaborative project QuProCS (Grant Agreement 641277). Moreover, we thank Thomas Wellens, Alberto Rodriguez, Rodolfo Jalabert, and Cord M\"uller for helpful comments on the manuscript.


\section{Appendix}

In the following we evaluate the short-time disorder dynamics (\ref{Eq:Short-time_master_equation}) for two simple, yet relevant examples: a particle of mass $m$ in one-dimensional, continuous space, subject to a random i) linear or ii) harmonic potential.

\paragraph*{i) In the linear-potential case,} we consider a Hamiltonian of the form $\hat{H}_{\varepsilon} = \hat{p}^2/2 m + \varepsilon \hat{x}$, i.e.~the randomness lies in the strength of the constant force exerted on the particle. This describes for example experiments where a charged particle is exposed to a homogeneous, but not fully controlled electric field, i.e.~the field strength varies from run to run. If we assume for simplicity that $\overline{\varepsilon} = 0$, the average Hamiltonian corresponds to the free Hamiltonian, $\hat{\overline{H}} = \hat{p}^2/2 m$, and for the Lindblad operators we obtain $\hat{L}_{\varepsilon} = \varepsilon \hat{x}/E_0$, with $E_0$ an arbitrary energy scale (which is again introduced for dimensional reasons and irrelevant for the final result (\ref{Eq:Caldeira-Leggett_equation})). Since all Lindblad operators are proportional to $\hat{x}$, we can perform the disorder integral in (\ref{Eq:Short-time_master_equation}) and obtain the simplified master equation
\begin{equation} \label{Eq:Caldeira-Leggett_equation}
\dot{\overline{\rho}} = -\frac{\rm i}{\hbar} \left[ \frac{\hat{p}^2}{2 m}, \overline{\rho} \right] - \frac{\overline{\varepsilon^2}}{\hbar^2} t \, [\hat{x}, [\hat{x}, \overline{\rho}]] .
\end{equation}
This is the well-known Caldeira-Leggett master equation \cite{Caldeira1983path, Diosi1993high}, which usually emerges in an open-system context from a linear coupling model. The incoherent part of (\ref{Eq:Caldeira-Leggett_equation}) predicts an exponential decay of spatial coherences according to (as for the derivation of the collisional decoherence localization function (\ref{Eq:Localization_function}), we neglect for the moment the von Neumann commutator) $\overline{\rho}_t(x, x') = \exp \left[ - \frac{\overline{\varepsilon^2}}{2 \hbar^2} t^2 (x-x')^2 \right] \rho_0(x, x')$. While such a localization rate which grows above all bounds for $|x-x'| \rightarrow \infty$ is usually considered as unphysical in the open-system context, it arises here as a natural and unavoidable consequence of the disorder average.

\paragraph*{ii) In the harmonic-potential example,} we could in principle allow for both a random frequency and a random center point. We focus here on the latter and keep the frequency fixed, $\hat{H}_{\varepsilon} = \hat{p}^2/2 m + (m \omega^2/2) (\hat{x} - \varepsilon)^2$. This may describe experiments where a particle is harmonically trapped, but where the trap center is subject to fluctuations. In this case (again assuming that $\overline{\varepsilon} = 0$), the short-time dynamics (\ref{Eq:Short-time_master_equation}) result, once again, in Caldeira-Leggett decoherence, $\dot{\overline{\rho}} = - ({\rm i}/\hbar) [\hat{\overline{H}}, \overline{\rho}] - m \omega^2 (\overline{\varepsilon^2}/\hbar^2) t [\hat{x}, [\hat{x}, \overline{\rho}]]$, but this time with a harmonic average potential $\hat{\overline{H}} = \hat{p}^2/2 m + (m \omega^2/2) \hat{x}^2$. In this example, we can even anticipate the evolution of the ensemble average beyond the short-time approximation: Since all random potentials share the same frequency $\omega$, any initial state will at multiples of the period $T = 2 \pi/\omega$ recur, and in particular it will regain the purity lost in the early stage. On the level of the disorder master equation, this indicates periodic, partly negative decoherence rates $\gamma_{\varepsilon}(t)$. While such time dependence comprising (at least partial) purity revivals is likely the generic pattern of the ensemble average dynamics, Fig.~\ref{Fig:Master_equation_comparison}(d) indicates that the Anderson model exhibits a strictly monotonic decay of coherences, also beyond the short-time approximation.

\end{document}